\documentstyle[aps,pre,epsf,floats]{revtex}

\begin{document}
\draft
\twocolumn[\hsize\textwidth\columnwidth\hsize\csname@twocolumnfalse%
\endcsname

\title  {Flowing sand - a physical realization of 
	 Directed Percolation}

\author {Haye Hinrichsen$^1$, Andrea Jim\'enez-Dalmaroni$^{2,3}$, 
	Yadin Rozov$^2$, and Eytan Domany$^2$\\[2mm]}

\address{$^{1}$
        Max-Planck-Institut f\"ur Physik komplexer Systeme,
        N\"othnitzer Stra\ss e 38, 01187 Dresden, Germany \\       
 	$^{2}$ 
	Department of Physics of Complex Systems,
        Weizmann Institute of Science, Rehovot 76100, Israel\\
	$^{3}$ 
	University of Oxford, Department of Physics - 
	Theoretical Physics, 1 Keble Road, Oxford OX1 3NP, U.K.       
 }
\date   {October 25, 1999}
\maketitle

\begin{abstract}
We introduce and investigate a simple model to describe 
recent experiments by Douady and Daerr on flowing sand.
The model reproduces experimentally observed compact avalanches, whose
opening angle decreases linearly as a threshold is approached.
On large scales the model exhibits a crossover from compact directed
percolation to directed percolation; 
we predict similar behavior for the experimental system. We
estimate the regime where ``true'' directed 
percolation morphology and exponents 
will be observed, providing the first experimental 
realization for this class of models. 

\end{abstract}

]
%
%
%
%

Directed Percolation (DP) is perhaps 
the simplest model that exhibits a
non-equilibrium phase transition between an 
``active'' or ``wet'' phase and an
inactive ``dry'' one~\cite{DP}. 
In the latter phase the system is 
in a single ``absorbing'' state; once it reaches the
completely dry state, it will always stay there. 

Interest in DP stems mainly from  
{\it universality} of the associated
critical behavior. It is believed 
that transitions in all models
with an absorbing state are in the 
DP universality class (unless there are some
special underlying symmetries). 
Even though DP exponents have not yet been 
calculated analytically, their values
were measured for a wide variety of models and are
known (especially in 1+1 dimensions) 
with very high precision~\cite{Jensen}.

Models in the DP universality class 
are supposed to describe a wide variety
of phenomena, ranging from catalysis to 
turbulence~\cite{Pomeau}; nevertheless, so far 
no physical system has been found 
that exhibits DP behavior and exponents~\cite{Grassberger96}. 
We show here that a simple system of sand flow on an 
inclined plane, recently studied by Daerr and  
Douady (DD)~\cite{DouadyDaerr}, may well be the first 
physical realization of a transition in the DP universality 
class. 

The results reported by DD present a puzzle, namely
the threshold phenomenon they discovered exhibits ``wet" 
clusters whose shapes differ from those seen in 
standard DP simulations; they are much more
compact. The corresponding model, 
called Compact Directed Percolation
(CDP) is unstable against perturbations 
towards the standard DP behavior
\cite{DomanyKinzel};
the latter is the generic case 
expected to occur. Since DD did no 
fine-tuning to place their system in 
the CDP class, their observations are
surprising.  

This motivated us to look for a   
simple model, which is defined in terms of 
dynamic rules that can be plausibly 
related to the experiments and  
exhibits features that reproduce the   
experimentally observed ones. We then investigated 
whether the transition exhibited by such a model 
belongs to the DP universality 
class. 

We propose a {\it directed sandpile} 
model, simpler than that of 
Tadic and Dhar~\cite{TadicDhar97}; 
here after each avalanche the system is
reset to a uniform initial state.  
Our model has a transition from an 
inactive to an active phase, in which we see 
avalanches whose compact shapes  
reproduce the experimental observations 
and indeed, the resulting critical 
behavior is close to CDP, rather than to DP.
We resolve this by showing that the CDP 
type critical behavior is a transient:
the true critical behavior {\it is} of the DP type, 
but it can be seen only after a very long
crossover regime. Our conclusion is that the DD 
experiment does serve as a possible
realization of a DP-type transition. 
We propose here ways to shorten the crossover regime
and to extend the scale on which the experiments are performed.

\vspace{2mm}\paragraph{The Douady-Daerr experiment. \ }

Glass beads (``sand'') of diameter 
$250$-$425 \ \mu$m~\cite{DouadyDaerr} 
are poured uniformly at the top of
an inclined plane (size $\sim$
$1m$), covered by a rough velvet cloth; 
the angle of inclination  
$\varphi _{0}$ can
be varied.  As the beads flow down, 
a thin layer of thickness   
$h=h_{d}(\varphi _{0})$, consisting 
of several monolayers, settles and
remains immobile. At this thickness the sand 
is {\it dynamically stable}; the value 
of $h_d$ decreases with increasing 
angle of inclination. 

For each $\varphi_0$ there exists another thickness $h_s$ with 
$h_s(\varphi _{0}) > h_{d}(\varphi _{0})$, 
beyond which a {\it static} layer
becomes unstable. Hence there exists a region in the 
$(\varphi,h)$ plane, in which a static 
layer is stable but a flowing one is
unstable. We can now take the system, 
that settled at $h_{d}(\varphi _{0})$, 
and increase its angle of inclination
to $\varphi=\varphi_0+\Delta \varphi$, staying within this 
region of mixed stability. The layer will 
not flow spontaneously, but if we 
disturb it,  generating a flow at the top, 
an avalanche will propagate, leaving 
behind a layer of thickness
$h_d(\varphi)$. These avalanches 
had the shape of a fairly regular triangle
with opening angle~$\theta$.
As the increment $\Delta \varphi$
decreases, the value of $\theta$ 
decreases as well, vanishing as 
$\Delta \varphi \rightarrow 0$. 
This calls for testing a power law behavior
of the form
\begin{equation}
\theta \sim (\Delta \varphi)^x \,.
\label{eq:powerlaw}
\end{equation}
If instead of increasing $\varphi$ 
we lower the plane, i.e.,  go to
$\Delta \varphi <0$, the thickness of our system, 
$h_{d}(\varphi _{0})$,
is less than the present thickness of 
dynamic stability, $h_{d}(\varphi )$.
In this case an initial perturbation should not 
propagate, it will rather die out 
after a certain time (or beyond a certain size 
$\xi_\parallel$ of the transient avalanche). As  
$ \vert \Delta \varphi \vert \rightarrow 0$, 
we expect  
this decay length to grow with a power law:
\begin{equation}
\xi_\parallel \sim (-\Delta \varphi)^{-\nu_\parallel} \,.
\label{eq:powerlaw2}
\end{equation}
Hence by pouring sand at inclination 
$\varphi_0$, DD  produced a critical 
system, precisely at the borderline
(with respect to changing the angle) 
between a stable regime $\varphi < \varphi_0$, 
in which perturbations die out, and an unstable one 
$\varphi > \varphi_0$, where perturbations persist and spread.
The preparation procedure can be considered as a special kind of 
{\it self-organized criticality} (SOC)
which differs from standard
SOC models~\cite{Bak} in which a slow 
driving force (acting on a time scale
much smaller than that of the system's 
dynamic response) causes evolution to 
a critical state. Here
avalanches are started by hand one by one. 

To associate this threshold phenomenon with
DP, denote by~$p$ the percolation probability
and by $p_c$ its critical value. 
We associate the change in tilt with $p - p_c$, i.e., 
assume that near the angle of preparation $\varphi_0$
the behavior of the sand system 
is related to a DP problem with
$ \Delta \varphi  \propto p - p_c $.
The exponent $\nu_\parallel$ should be 
compared with the known values for
DP and CDP. The exponent $x$ in Eq.~(\ref{eq:powerlaw}) 
can also be measured and compared with 
\begin{equation}
\tan \theta \sim \xi_\perp / \xi_\parallel \sim 
(\Delta \varphi )^{\nu_\parallel - \nu_\perp } \,.
\label{eq:Dnu}
\end{equation}
%
%

\vspace{2mm}\paragraph{Definition of the model.}

\begin{figure}
\epsfxsize=70mm
\centerline{\epsffile{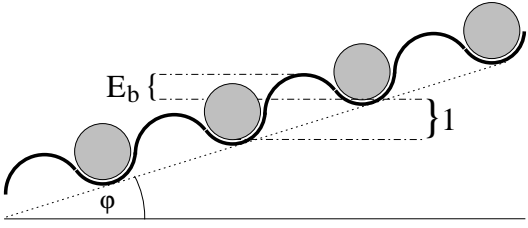}}
\vspace{2mm}
\caption{
\label{FigPotential}
The top layer of sand in a washboard potential.
}
\end{figure}

To write down a simple model based on the physics of 
the flowing sand, we adopt an observation 
made by DD, that in the regime of interest 
($\varphi \approx \varphi_0$) grains of the top layer 
rest on grains of the layers below (rather 
than on other grains of the top layer). 
Hence the lower layers provide for
the top one a  {\it washboard potential}, 
as shown in Fig.~\ref{FigPotential}.

We place the grains of the top layer 
on the sites of a regular square 
lattice with row index $t$ and columns $i$. 
At any given time a grain $G$ may become active 
if at least one of its neighbors from 
the row above has been active at the previous 
time step. If $\Delta E(G)$, the total energy 
transferred from these neighbors,
exceeds the barrier $E_b$ of the washboard, 
$G$ becomes active, ``rolls down'' and 
collides with the grains of the next row. 
The energy it brings to  these collisions
is $1~+~\Delta E(G)$, where 1 is 
our unit of energy, representing the
potential energy due to the height 
difference between two consecutive rows
(see Fig.~\ref{FigPotential}).
A fraction $f$ of its total energy is
dissipated; the rest is divided stochastically 
among its three neighbors from the lower row. 

The model is defined in terms of 
two variables; an activation variable
$S^t_i = 0,1$ and an energy $E^t_i$. 
The index~$t$ denotes rows of our
square lattice {\it and} time; 
at time~$t$ we update the states of the grains
that belong to row~$t$. The model is controlled by two
parameters: $E_b$, the barrier height, and $f$,
the fraction of dissipated energy. 

The dynamic rules of our model are 
as follows. For given  activities $S_i^{t}$ and 
energies $E_i^{t}$  we first calculate the energy transferred 
to the grains of the next row $t+1$.
To this end we generate for each active site three random
numbers, $z_i^t(\delta )$ (with $\delta = \pm1,0$) that add up to 1.
The energy transferred to grain $(t+1,i)$, given by
\begin{equation}
\Delta E_i^{t+1}~~=~~(1-f)\sum_{\delta=\pm 1,0} S^{t}_{i-\delta} 
~ E^{t}_{i-\delta}
~ z_{i-\delta}^{t}(\delta)  \,,
\label{eq:DEt}
\end{equation}
determines its  activation: 
\begin{equation}
S^{t+1}_i= \left\{
\begin{array}{cc}
1   &   \mbox{active $~~$ if $~~~\Delta E^{t+1}_i > E_b$} \,, \\
0   &   \mbox{inactive if $~~~\Delta E^{t+1}_i \leq E_b$} \,.
\end{array}
\right.
\label{eq:St}
\end{equation}
Then the energies of the next row of grains are set:
\begin{equation}
E_i^{t+1}~ = ~ S_i^{t+1} ~ (1 + \Delta E_i^{t+1}) \,.
\label{eq:Et}
\end{equation}
The three 
random numbers $z_i^{t}(\delta)$  represent 
the fraction of energy transferred  from  the grain 
at site $(t,i)$ to the one at $(t+1,i+\delta)$. 
We add up the energy contributions from these 
active sites; the fraction $1-f$ is {\it not}
dissipated; if the acquired energy $\Delta E_i^{t+1}$ exceeds $E_b$, 
site $(t+1,i)$ becomes active, rolls over the barrier and 
brings to the collisions (at time $t+2$) the acquired energy 
calculated above {\it and} its excess potential energy (of value 1).

\vspace{3mm}\paragraph{Qualitative discussion of the transition 
and connection to the experiment.}

\begin{figure}
\epsfxsize=75mm
\centerline{\epsffile{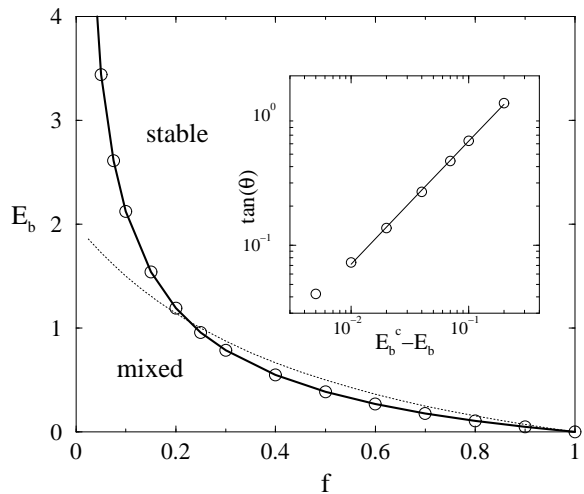}}
\caption{
\label{FigPhaseTrans}
Phase diagram of the model for flowing sand. 
The full line represents the phase transition 
line. The mean field approximation
of Eq.~(\ref{SimpleMeanField}) is
shown as a dotted line. The inset shows the 
opening angle $\tan{\theta}$ as a function of $E_b^c-E_b$.
}
\end{figure}

Let us  vary $E_b$ at a 
fixed value of the dissipation. For small values of 
$E_b$ an active grain will activate 
the grains below it with high probability;
avalanches will propagate downhill and also spread sideways. For a 
strongly localized initial activation we should, therefore, see 
triangular shaped activated regions. As $E_b$ increases,
the rate of activation decreases and the opening angle 
$\theta$ of these triangles should decrease, 
until $E_b$ reaches a critical value $E_b^c$, beyond which initial
activations die out in a finite number of time steps (or rows).
These expectations are indeed borne out by simulations of the model:  
the dependence of $E_b^c$  on the dissipation $f$ 
is shown in Fig.~\ref{FigPhaseTrans}.

The physics of the process is captured by a
simple mean-field type approximation, in which all stochastic
variables are replaced by their average values.
Consider an edge separating an active
region from an inactive one. At time $t$ sites
to the left of $i$ and $i$ itself are wet, 
whereas $i+1,i+2,...$ are dry.  
Will  site $i$ be wet or dry 
at the next time step? In our mean-field
estimate of the answer, assuming that all wet 
sites at time $t$ have the same energy $E^t$,
the energy delivered to site $i$ at time $t+1$ is
$
\Delta E^{t+1}_i=\frac{2}{3}(1-f)(1+\Delta E^{t})$,
where we set in Eq.~(\ref{eq:DEt})
all $z(\delta)=1/3$. At the critical point we expect all energies 
to just suffice to go over the barrier; hence set  
$\Delta E_i^{t+1}=\Delta E^t=E_b^c$.  
Solving the resulting equation yields a
rough estimate of the 
transition line,
\begin{equation}
E_b^c=2(1-f)/(1+2f) \, ,
\label{SimpleMeanField}
\end{equation}
as shown in Fig.~\ref{FigPhaseTrans}. 
It is easy to produce better mean-field 
type estimates of the transition and 
to compute the corresponding energy profile  
of the wet region~\cite{Future}.

%
%

To connect our model to the DD
experiments note that
the tilt angle $\varphi$ 
tunes the ratio between
the barrier height and the difference 
of potential energies between two rows.
When the system is prepared at $\varphi_0$,
this ratio is precisely~$E_b^c$. 
When one increases the tilt angle to $\varphi > \varphi_0$, 
$E_b$ (measured in units of the 
potential difference) decreases  
and we have $E_b < E_b^c$. 
As the tilt angle is now reduced, 
the  size of $E_b$   increases, 
until it  reaches its critical value 
precisely at $\varphi_0$. Thus increasing 
$E_b$ in the model corresponds to 
lowering the tilt angle towards $\varphi_0$ where 
the system is precisely at its boundary of dynamic stability.

Hence to reproduce the experiment we were looking for 
(a) fairly compact triangular regions of activation 
for $E_b <E_b^c$, and (b) an opening angle of these triangles 
which should go to zero as $E_b$ approaches $E_b^c$ from below.

\begin{figure}
\epsfxsize=85mm
\centerline{\epsffile{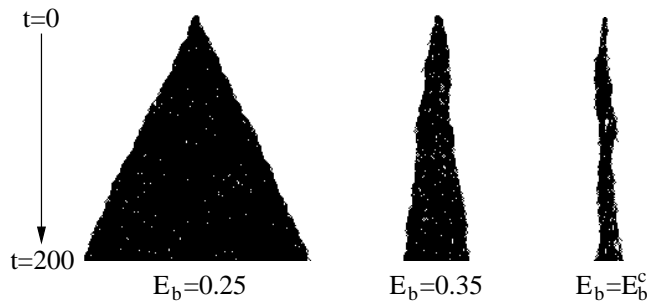}}
\vspace{2mm}
\caption{
\label{fig:avalanches}
Typical avalanches starting from a single seed
with dissipation $f=0.5$ far away and close
to criticality.
}
\end{figure}

We simulated the model defined 
in Eqs.~(\ref{eq:DEt})-(\ref{eq:Et}) and found that it indeed 
reproduces 
these qualitative features of the experiment (see 
Fig.~\ref{fig:avalanches}).
The avalanches shown were produced for dissipation 
$f=0.5$, activating a single site at 
$t=0$, to which an initial energy of $E_0=100$ was assigned.
As long as $E_b$ was not too close to $E_b^c$
the observed avalanches were compact, triangular and with fairly straight 
edges. The edges became rough only very close to $E_b^c$, such as
the one shown on the right
hand side of Fig.~\ref{fig:avalanches}. 
The opening angle of the active regions $\theta$ decreased as 
$E_ b$ increased towards $E_b^c$, as indicated in 
in the inset of Fig.~\ref{FigPhaseTrans}. From these simulations we 
estimated $E_b^c$ and the exponent (see Eq.~(\ref{eq:Dnu}))
\begin{equation}
x=\nu_\parallel - \nu_\perp = 0.98(5) \simeq 1 \,.
\label{linear}
\end{equation}

The linear variation of $\tan(\theta)$ with $\Delta \varphi$
is in agreement with experimental measurements~\cite{DouadyDaerr}.
Our findings have to be compared with the mean-field theory suggested
in Ref.~\cite{BouchaudCates98} which predicts a square root behavior.

\vspace{3mm}\paragraph{Crossover to directed percolation.}

The linear law (\ref{linear}) 
is consistent with the critical 
exponents of CDP~\cite{DickmanTretyakov95}
\begin{equation}
\nu_\parallel=2\,, \qquad \nu_\perp=1 \,, \qquad \beta=0\,.
\end{equation}
These observations pose, however, a 
puzzle: since one believes that DP
is the generic situation, one would 
expect to find non-compact active regions
and DP exponents. In fact, according 
to the DP conjecture~\cite{DPConjecture}
any continuous spreading transition 
from a {\em fluctuating} active phase
into a single frozen state should 
belong to the universality class of
directed percolation (DP), provided 
that the model is defined by short
range interactions without exceptional 
properties such as higher
symmetries or quenched randomness.
The present model has neither special 
symmetries nor randomness~\cite{random}; 
it  has a fluctuating active phase and exhibits a
transition, characterized by a 
positive one-component order parameter, 
into a single absorbing state. 
Hence the phase transition of our model should
belong to the DP universality class.
%
%
%
\begin{figure}
\epsfxsize=80mm
\centerline{\epsffile{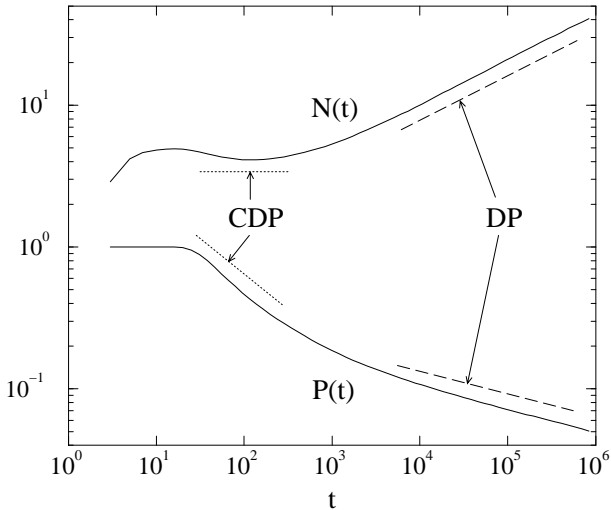}}
\caption{
Average number of active sites $N(t)$ and
mean survival probability $P(t)$ of the toppling process
at criticality averaged over $50\,000$ runs.
The predicted slopes for CDP and DP are indicated by
dotted and dashed lines, respectively.
\label{FigSurv}
}
\end{figure}

In order to understand this apparent paradox we
performed high-precision Monte-Carlo simulations for
dissipation $f=0.5$ (see~\cite{Future} for further details). 
We performed time-dependent simulations~\cite{GrassbergerTorre79},
i.e., we toppled a single grain in the center of the top row and
measured the survival probability $P(t)$ and 
the number of active sites $N(t)$.  
At criticality, these quantities exhibit
an asymptotic power law behavior
\begin{equation}
P(t)\sim t^{-\delta}\,,
\qquad
N(t) \sim t^\eta\,.
\end{equation}
In the case of CDP these
exponents are given by~\cite{DomanyKinzel,DickmanTretyakov95}
$\delta=1/2$ and $\eta=0$, 
whereas DP is characterized by the exponents~\cite{Jensen}
$\delta=0.1595$ and $\eta=0.3137$.
Detecting deviations from power-law behavior in the long-time limit
we estimated $E_b^c=0.385997(5)$.

Numerical results, obtained from simulations at $E_b^c$, 
are shown in Fig.~\ref{FigSurv}. After a 
short transient the system
enters an intermediate regime, which extends up to several
hundred time steps. Here the active sites form a single cluster and 
we observe power-law behavior with CDP exponents
(dotted lines in Fig.~\ref{FigSurv}).
This intermediate regime is followed by a long
crossover from CDP to DP, extending over almost two decades up to
more than $10^4$ time steps, after which the system enters an 
asymptotic DP regime (indicated by dashed lines 
in Fig.~\ref{FigSurv}).

Compared with ordinary DP lattice models, this cross\-over 
regime is extremely long. We observed that by 
increasing $f$ the crossover time can be reduced by 
more than one decade. Hence, for an experimental 
verification of DP, systems with high 
dissipation are more appropriate. The present experiments
correspond to about 3000 time steps (rows of beads);
increasing this to about $10^4$ by using a longer inclined 
plane and smaller beads should yield DP behavior, provided 
that deviations of the experiment from the model do not 
increase with system size.
 
\begin{figure}
\epsfxsize=80mm
\centerline{\epsffile{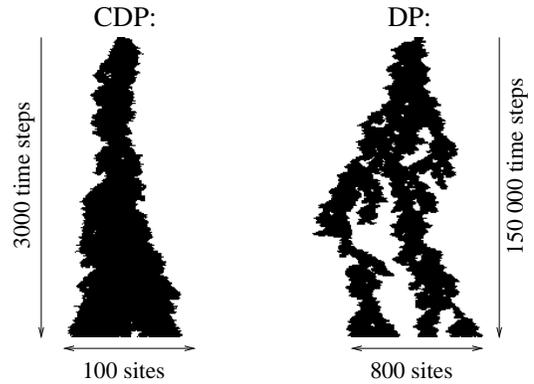}}
\vspace{2mm}
\caption{
Typical clusters generated at 
criticality on small and large scales,
illustrating the crossover from CDP to DP.
\label{FigDemoCrossover}
}
\end{figure}

The crossover from CDP to DP is illustrated in
Fig.~\ref{FigDemoCrossover}.
Two avalanches are plotted on different scales. The 
left one represents a typical avalanche within the first
few thousand time steps. As can be seen, the
cluster appears to be compact on 
a lateral scale up to 100 lattice sites.
However, as can be seen in the 
right panel of Fig.~\ref{FigDemoCrossover},
after very long time the cluster 
breaks up into several branches,
displaying the typical patterns of critical DP clusters.
Thus, before measuring critical exponents, this feature 
has to be tested experimentally.
To this end the DD experiment should 
be performed repeatedly at the critical tilt 
$\varphi=\varphi_0$. In most cases the avalanches will be
small and compact. However, sometimes large avalanches
will be generated which reach the bottom of the plate.
If these avalanches are non-compact,
we expect DP-type asymptotic critical
behavior. Only then is it
worthwhile to optimize the experimental setup and
to measure the critical exponents quantitatively.

%
%
%

AJD thanks the CICPB and the UNLP,
Argentina and the Weizmann Institute for financial support.
HH thanks the Weizmann Institute and the 
Einstein Center for hospitality and financial support.
ED thanks the Germany-Israel Science Foundation for 
support, B. Derrida for some most helpful initial insights
and A. Daerr for communicating his results at an early stage.


\end{document}